%% file: main.tex
\documentclass[conference]{IEEEtran}

\IEEEoverridecommandlockouts
\usepackage{cite}
\usepackage[inline]{enumitem}

\usepackage{cuted}

\usepackage{booktabs} 

\usepackage{graphicx}
\usepackage[justification=centering]{caption}
\usepackage{wrapfig}
\usepackage{subcaption}

\usepackage{color}

\usepackage{amsmath}
\usepackage{multirow}
\usepackage{array}
\newcolumntype{P}[1]{>{\centering\arraybackslash}p{#1}}

\usepackage[table]{xcolor}
\definecolor{blue_t}{RGB}{221,221,255}
\newcommand{\bc}[1]{\cellcolor{blue_t}}

\definecolor{red_t}{RGB}{255,221,221}

\definecolor{green_t}{RGB}{169,208,142}
\newcommand{\gc}[1]{\cellcolor{green_t}}

\definecolor{peach_t}{RGB}{244,176,132}
\newcommand{\pc}[1]{\cellcolor{peach_t}}

\definecolor{grey_t}{RGB}{201,201,201}
\newcommand{\grc}[1]{\cellcolor{grey_t}}

\definecolor{yellow_t}{RGB}{255,192,0}
\newcommand{\yc}[1]{\cellcolor{yellow_t}}
\usepackage{nicematrix}

\usepackage{lipsum}

\begin{document}

\bstctlcite{IEEEexample:BSTcontrol}

\title{Design and Optimization of  Magnetic-core Solenoid  Inductor for Multi-phase Buck Converter \vspace{-0.5cm}}

\author{ \IEEEauthorblockN{Madhava Sarma Vemuri} \IEEEauthorblockA{ Department of Electrical and Computer Engineering\\
  North Dakota State University\\
  Fargo, North Dakota 58105 \\
  }
\and
\IEEEauthorblockN{Umamaheswara Rao Tida} \IEEEauthorblockA{ Department of Electrical and Computer Engineering\\
  North Dakota State University\\
  Fargo, North Dakota 58105 \\
  }
\thanks{Madhava Sarma Vemuri is with the Department of Electrical and Computer Engineering, North Dakota State University, North Dakota, United States -- 58102}
\thanks{Umamaheswara Rao Tida is with the Department of Electrical and Computer Engineering, North Dakota State University, North Dakota, United States -- 58102. e-mail: umamaheswara.tida@ndsu.edu}
\thanks{Manuscript received: Oct 14, 2019;}
\vspace{-1cm}}

\title{FDSOI Process Based MIV-transistor Utilization for Standard Cell Designs in Monolithic 3D Integration
(Accepted in SOCC 2023)} 

\maketitle

\begin{abstract}
Monolithic Three-Dimensional Integrated Circuits (M3D-IC) has become an attractive option to increase the transistor density. In M3D-IC, substrate layers are realized on top of previous layers using sequential integration techniques. Recent works in M3D-IC have demonstrated the feasibility of FDSOI process-based M3D-IC implementations and, Metal inter-layer vias (MIVs) are used to provide connections between the inter-layer devices. Since MIVs are extended from bottom layer to top layer, they occupy a small area resulting in area overhead. Additionally, a minimum separation is required to facilitate connection between MIV and transistors which increases this overhead further. Towards this, we studied the alternate utilization of MIV to create MIV-transistors with varying channels. We have also presented a strategy to extract the Spice parameters of the proposed models using level 70 spice parameters. Finally, a standard cell based gate level comparison is presented to compare the Power, Performance and Area (PPA) metrics of the traditional two layer 2D FDSOI transistor implementation with the proposed models. Simulation results from standard cell designs suggest that the proposed methodology can reduce 18\% layout area on average compared to the traditional approach. In addition, power consumption and delay time of the standard cells are reduced by 1\% and 3\% on average respectively.

\end{abstract}


\input{section1_intro}

\input{Section2_background}
\section{FDSOI Process}
\input{Section2_FDSOI_process}\label{sec:FDSOI_Process}


\section{Modeling and Design of MIV-transistor in FDSOI-Process}
\input{Section4_MIVFET}\label{sec:MIVFET}

\subsection{Spice Extraction Initial Setup}
\input{Section4a_TCAD2SPICE}\label{sec:TCAD2SPICE}

\subsection{Extraction Results}\label{subsec:extraction_results}
\input{Section4b_Extraction_results}

\section{Power, Performance and Area comparison of Standard cells }\label{sec:ppa_comparison}
\input{Section5_Standard_cell_simulation_results}






\section{Conclusions}\label{sec:conclusion}
\input{section6_Conclusions}

\section*{Acknowledgements}
This work is supported by National Science Foundation under Award number -- 2105164.

\bibliographystyle{IEEEtran}
\bibliography{references}

\end{document}

%% file: Section1_intro.tex
\section{Introduction}\label{sec:introduction}
In recent years, vertical integration has piqued interest in the IC design industry. Vertical integration allows stacking active devices on top of each other. Consequently, it improved the transistor density, and reduced the global interconnection wire length, making it a viable alternative to conventional transistor scaling. Previously, Vertical integration  techniques stacked processed substrate layers and, used Through-Silicon-Vias (TSVs) to provide connections between processed layers. However, the size of TSV is limited by the substrate die thickness, and does not scale with transistor scaling \cite{tida_resonant_clocking_tvlsi,tida_efficacy_tvlsi}. Aggressive scaling of TSV size has only resulted in TSV diameters $>2 \mu m$, which is still larger than conventional standard cell dimension \cite{sklim_14nm}. 

To address the limitations of TSV-based scaling techniques, Monolithic three Dimensional Integrated circuit (M3D-IC) technology has been proposed.  M3D-ICs are realized by sequential integration of active layers, which reduced the substrate thickness to $<0.1 \mu m$. Existing works on  M3D-IC process such as CoolCube\textsuperscript{TM} technology sets a thermal budget and limit the process temperature to $< 500^oC$ \cite{thermal_budget_500,thermal_budget_500_2,thermal_budget_450_2}. The thermal budget prevents the degradation of transistors in the subsequent bottom layers. An Inter-layer-Dielectric (ILD) region is created to reduce the interactions between top and bottom layers. To facilitate the interconnections between the top and bottom layers, Metal Inter-Layer Via (MIV) is used. A thin ILD region allows increased MIV density due to reduced Via sizes ( $< 0.1 \mu m$). For 14nm design rules, the estimated  MIV density is $> 100$ M/mm\textsuperscript{2} \cite{MIV_integration_density}. Since MIVs are extended from bottom layer to top layer, they occupy a small area resulting in area overhead. Additionally, a minimum separation is required to facilitate connection between MIV and transistors which increases this overhead further.  


M3D-IC designs can be implemented with multiple abstraction styles such as block-level, gate-level, and transistor-level \cite{panth2013high_block_level_design,panth2014_gate_level_design,  rc_parasitics_M3DIC}. In block-level style, the functional blocks of standard cells are grouped, and placed in multiple layers. In gate-level abstraction style standard cells are placed in multiple layers, and the interconnects are routed. In the transistor-level style, the transistors are partitioned into different layers. In comparison to the previous design style, the transistor-level design allows finer integration, since partitioning is done at a granular level. In an ideal situation, M3D-IC offers 50\% area reduction due to the 2-layer partitioning of devices. However, only 40\% area reduction is reported for transistor-level design style, due to mismatch of n-type and p-type sizes and MIV area overhead in the top layer \cite{rc_parasitics_M3DIC}. 

Due to increasing MIV density and fine integration in transistor-level abstraction style, the MIV overhead also increased significantly. To address the MIV area overhead, alternate utilization has been investigated recently. Using the MIV-transistor, the silicon overhead was reduced by 24\% \cite{madhava2020MWSCAS,tida2020socc}. However, the study considers the entire top substrate layer to be silicon unlike the FDSOI Process used in the recent demonstrations \cite{vemuri_isqed}. 
 
 The major contributions of the work are as follows: 
\begin{enumerate}
    \item We have proposed FDSOI-based MIV-transistor models with varying channels, including 1-channel, 2-channel and 4-channel models.
    \item For comparison purposes, we have extracted the Spice parameters from TCAD simulations of conventional FDSOI 2D transistor and proposed MIV-transistors, using level 70 (BSIMSOI4) Spice parameters. 
    \item We have created 14 standard cells for Power, Performance and Area (PPA) comparisons. Our results suggest that, standard cells created using 2-channel MIV-transistors had shown a $3\%$ reduction in the overall power-delay-product and 18\%  average layout area reduction compared to the traditional 2-layer implementation.
\end{enumerate}

The organization of the rest of the paper is as follows: Section \ref{sec:FDSOI_Process} discusses in detail about the 2-layer M3D-IC in FDSOI process. Section \ref{sec:MIVFET} showcases multiple MIV-transistor models, and presents strategies for Spice parameter extraction of the models. Section \ref{sec:ppa_comparison} presents the PPA comparison and their implication of the standard cells created using the extracted Spice transistor model files. Finally, the concluding remarks are given in Section \ref{sec:conclusion}.

%% file: Section2_FDSOI_process.tex
\begin{figure}[]
    \centering
    \includegraphics[width=0.75\linewidth]{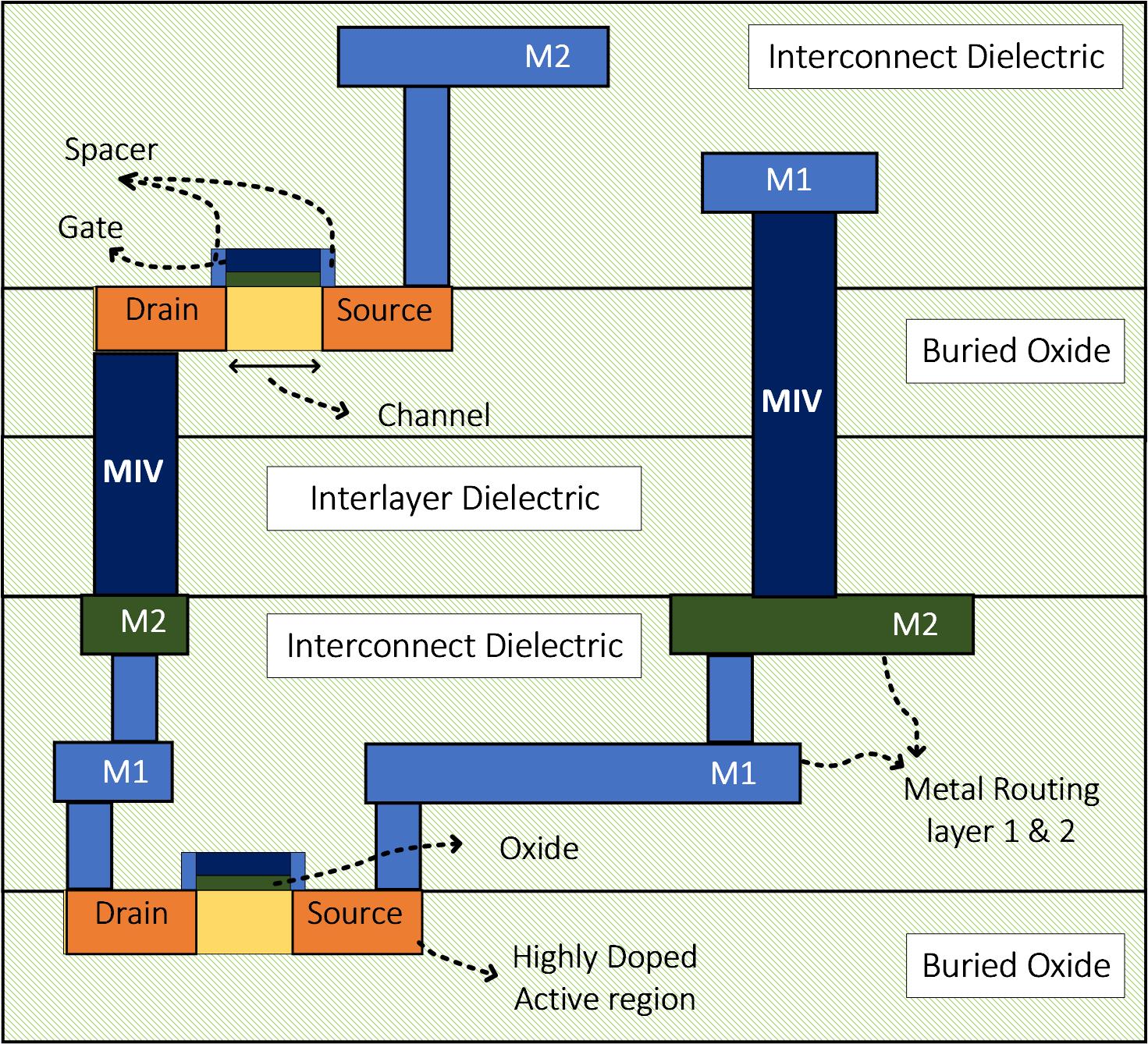}
    \caption{FDSOI Process in M3D-IC}
    \label{fig:FDSOI_process}
\end{figure}

In this section, we discuss about the assumed FDSOI process in M3D design throughout the paper. We have assumed a 2-layer M3D design where top and bottom layers are assigned for n-type and p-type transistors respectively. M3D-IC process with CoolCube\textsuperscript{TM} technology allows heterogeneous integration where the bottom layer transistors can be Bulk, FDSOI, FinFET or Trigate but the top layer transistors are realized only with FDSOI technology \cite{3dvlsi_pdsoi}. For simplicity, we have assumed both the top and bottom layers are created using FDSOI technology. The transistors in FDSOI technology are implemented over a dielectric region known as BOX region (\textbf{B}uried-\textbf{Ox}ide) \cite{UTBB_FDSOI}.  To facilitate the interconnections between the top and bottom layers, Metal-Interlayer-Vias (MIV) are used.  MIV is used as an, \begin{enumerate*} \item Internal Contact \item External Contact
\end{enumerate*} as shown in Figure \ref{fig:FDSOI_process}. When MIV is used as internal contact, Drain or Source active regions is be connected internally, and no additional area overhead is created. However, when MIV used as external contact to connect Gate region, it creates area overhead in the top substrate layer due to the presence of MIV. Additionally, minimum separation between the MIV and Gate region increases the area overhead. 

The illustration of the assumed process is shown in Figure \ref{fig:FDSOI_process}. A thin film undoped Silicon (Si) substrate (thickness of 7 nm) is used in the design of transistors. The p-type and n-type active regions are modeled using the Boron (B) and Arsenic (As) respectively. In this work, we assumed the active Source and Drain substrate region is doped using a concentration of $n_{src}$. The length and width of active regions are denoted by $l_{src}$ and  $w_{src}$ respectively. The length of the gate region ($L_{G}$) is assumed to be 24nm. We have used Silicon Nitride (Si\textsubscript{3}N\textsubscript{4}) to model the spacer region. All the insulator regions including the Buried Oxide (BOX), Interlayer Dielectric (ILD) and Interconnect Dielectric (ID) regions are modeled using Silicon dioxide (SiO\textsubscript{2}). All the interconnect routing layers (M1 and M2), MIV and Gate regions are modeled using Copper (Cu). The nominal value of thickness of MIV $t_{MIV}$ is assumed to be $25nm$. The width and thickness of M1 and M2 routing layer is assumed to be $24$nm and $48nm$ respectively, based on the 7nm PDK assumptions given in \cite{finfet_7nm_monolithic} for estimating the parasitics of interconnects. The size of via contact is assumed to be $24nm$. To model the device behavior in FDSOI process, we have created the CAD models using the Sentarus TCAD tool using the nominal values presented in Table \ref{tab:process_spec}. Shockley–Read–Hall (SRH) recombination model along with Fermi based statistics and Poisson equations are used to model the carrier behavior.

\begin{table}[htp]
\begin{center}
\caption{Process and design parameters used in the study}\label{tab:process_spec}
\begin{tabular}{|c|c|p{3.5cm}|c|}
    \hline
    & \textbf{Parameter} &  \textbf{Description} & \textbf{Value}\\
    \hline
    \multirow{6}{*}{\rotatebox{90}{\textbf{Process}}}& 
    $t_{Si}$ & Silicon Thickness  & $7$ nm \\
    \cline{2-4}
    & $h_{src}$ & Height of source/drain region & $7$ nm \\
    \cline{2-4}
    & $t_{ox}$ & Thickness of oxide liner & $1$ nm \\
    \cline{2-4}
    & $n_{src}$ & Source/Drain doping & $1\times10^{19}$cm\textsuperscript{-3} \\
    \cline{2-4}
    & $t_{spacer}$ & Spacer Thickness & $10$ nm \\ 
    \cline{2-4}
    & $t_{BOX}$ & Buried Oxide Thickness & $100$ nm \\ 
    \hline
    \multirow{3}{*}{\rotatebox{90}{\textbf{Design}}}& 
    t\textsubscript{miv} & MIV thickness & $25$ nm \\
    \cline{2-4}
    & $l_{src}$ & Length of Source/Drain region & $48$ nm \\
    \cline{2-4}
    & $w_{src}$ & Width of Source/Drain region & $192$ nm \\
    \cline{2-4}
    & $L_{G}$ & Length of Gate & $24$ nm \\
    \hline
\end{tabular}
\end{center}
\end{table}

%% file: Section4_MIVFET.tex
\begin{figure}

\begin{subfigure}[t]{0.46\linewidth}
\centering
\includegraphics[width=\linewidth]{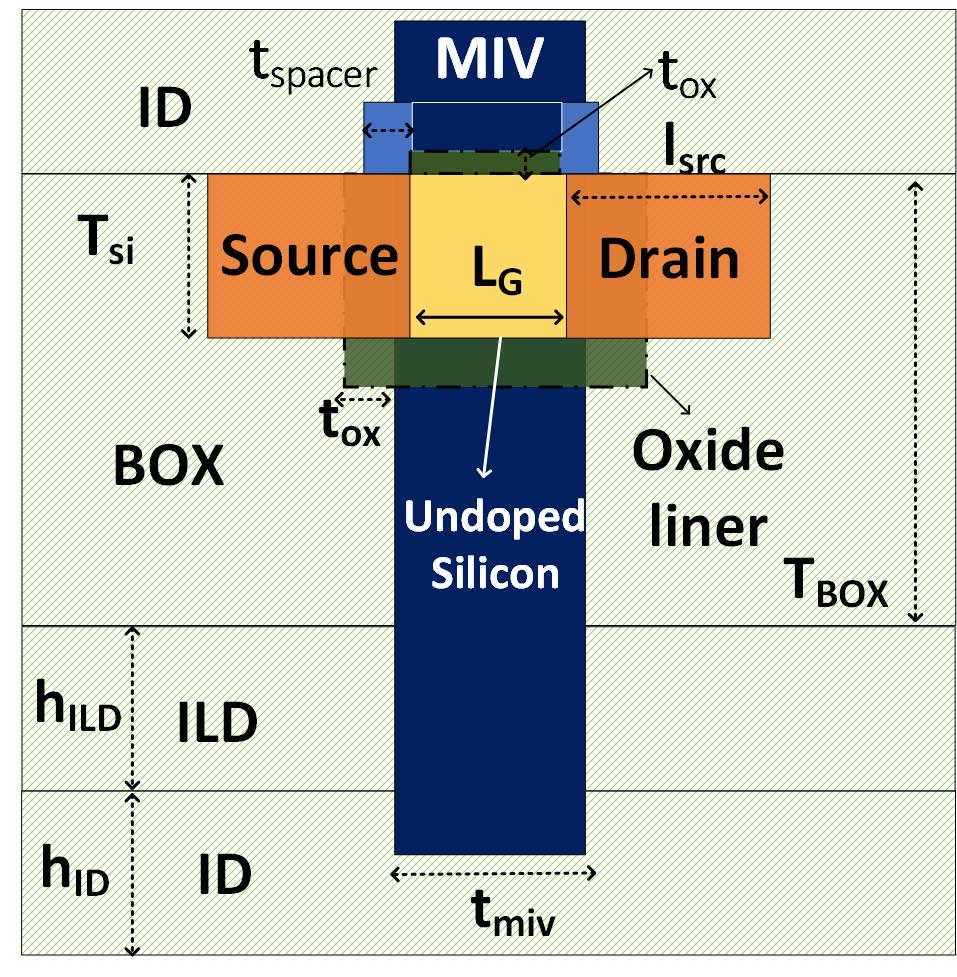}
    \caption{Side View}
    \label{subfig:mivfet_side_view}
\end{subfigure}
\begin{subfigure}[t]{0.46\linewidth}
\centering
\includegraphics[width=\linewidth]{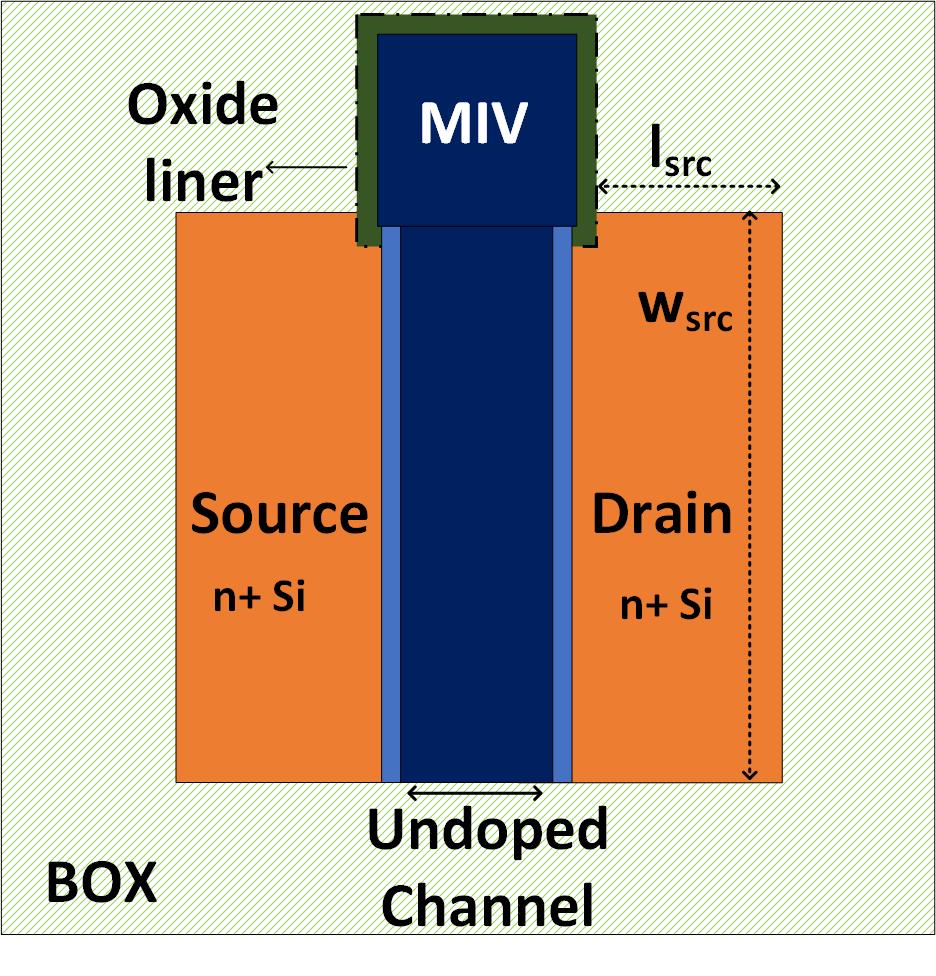}
    \caption{1-channel MIV-transistor Top View}
    \label{subfig:mivfet1c_top_view}
\end{subfigure}
\begin{subfigure}[t]{0.46\linewidth}
\centering
\includegraphics[width=0.96\linewidth]{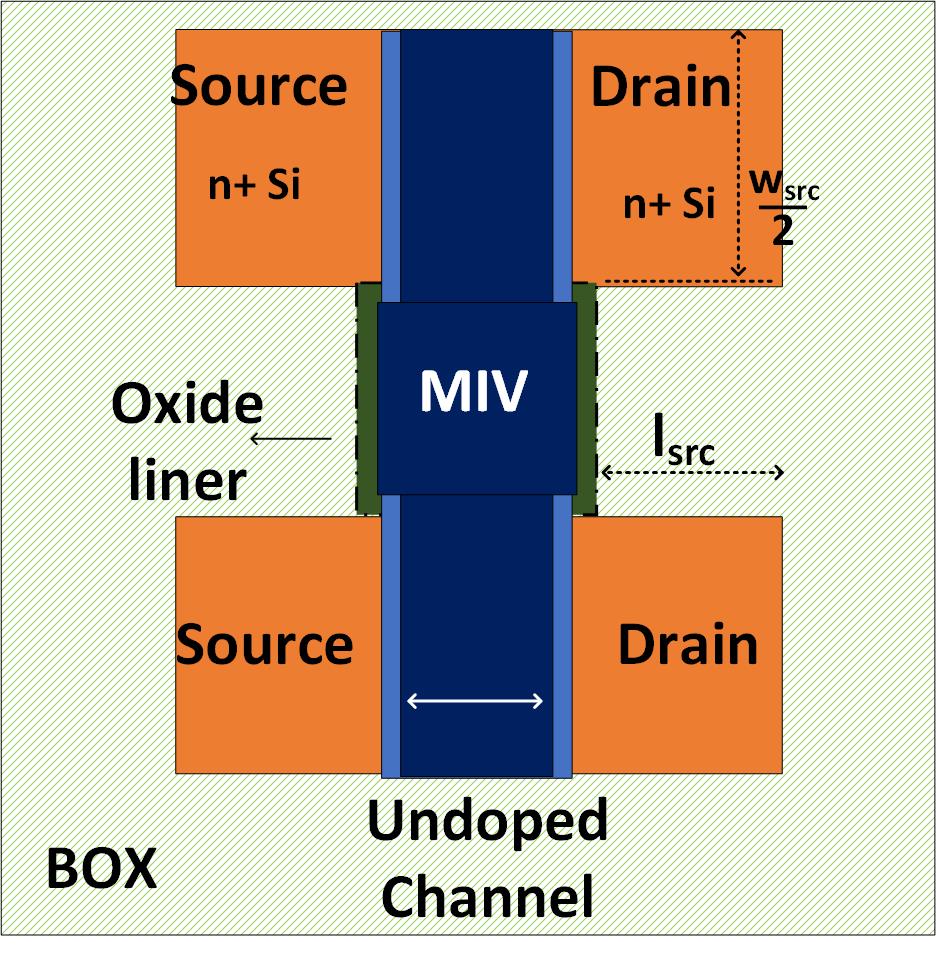}
    \caption{2-channel MIV-transistor Top View}
    \label{subfig:mivfet2c_top_view}
\end{subfigure}
\begin{subfigure}[t]{0.46\linewidth}
\centering
\includegraphics[width=\linewidth]{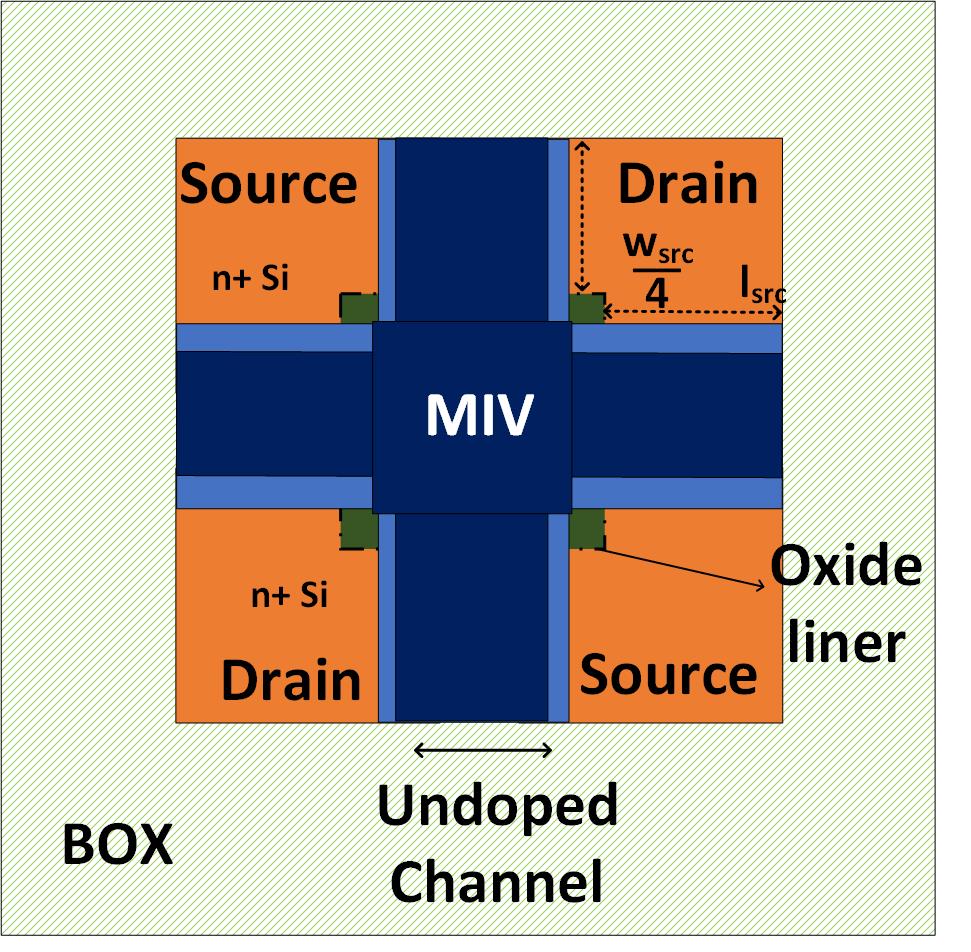}
    \caption{ 4-channel MIV-transistor Top View}
    \label{subfig:mivfet4c_top_view}
\end{subfigure}
\caption{Proposed layouts of MIV-based devices} \label{fig:MIVFET_models}
\end{figure}

As discussed earlier, the MIVs connecting to the gate induces extra area overhead and hence it needs to be addressed to realize efficient IC designs in M3D-IC technology. In this section, we discuss the modeling and design of MIV-transistor in M3D-IC technology, and spice extraction methodology in FDSOI process. The minimum thickness of oxide around MIV to provide electrical isolation to the substrate is assumed to be 1nm as shown in Figure \ref{subfig:mivfet_side_view}. The subtrate material is placed after this oxide liner to form metal-insulator-semiconductor (MIS) structure similar to MOS transistor \cite{madhava2020MWSCAS,tida2020socc}. To improve the channel control, the gate is realized on the top of the substrate similar to the conventional FDSOI implementation. This MIV-transistor structure can utilize substrate effectively reducing the MIV area overhead and also the wire length.

Various implementations of MIV-transistors specifically 1, 2 and 4-channel models are shown in Figure \ref{fig:MIVFET_models}. The 1-channel FDSOI technology based MIV-transistor model shown in Figure \ref{subfig:mivfet1c_top_view} has 1 Source and 1 Drain region similar to a traditional FDSOI transistor design. The Gate region and the MIV are connected together without any spacing between MIV and the gate terminal. However, the Source and Drain contacts should have minimum M1 spacing (which we assumed to be $24$ nm). Similarly,the 2-channel model has two Source and two Drain regions  as shown in Figure \ref{subfig:mivfet2c_top_view} where the Source and Drain regions should be connected together. Please note that for the basic gate designs, one metal layer is sufficient to provide necessary interconnects similar to the traditional FDSOI designs for 1-channel and 2-channel MIV-transistors. The 4-channel model has 2 Source and 2 Drain regions but has 4 channels as shown in Figure \ref{subfig:mivfet4c_top_view}. For 4-channel MIV transistor, the Source and Drain active regions are on the either side, making it complex for routing interconnects. Therefore, we need an additional routing  resources such as more interconnect wires to connect the Source and Drain regions respectively for the 4-channel MIV-transistor model.

The minimum dimension for the active region of 4-channel MIV-transistor is $48nm$, considering the smallest via size, and minimum separations between active region and gate contacts. For similar comparisons between the proposed MIV-transistor models, we assumed the width scaled exactly $2\times$ as we move from 4-channel model to 2-channel model, to account for the missing channel regions. Similarly, as we scale from 2-channel model to 1-channel model, the width scaled $2\times$. For fair comparison, we have assumed the equivalent width of transistor $w_{src}$ for all the models to be $192nm$. Therefore, the 1-channel Source and Drain region width will be $192nm$. The 2-channel Source and Drain regions width will be $96nm$ for each Source and Drain region shown in Figure \ref{subfig:mivfet2c_top_view}. Similarly, the 4-channel MIV-transistor active region width will be $48nm$ each for all active regions. As we can see from Figure \ref{subfig:mivfet4c_top_view}, there will be a total of four channels and hence the total width will be $196nm$.

%% file: Section4a_TCAD2SPICE.tex
TCAD modeling and design is not feasible for large designs due to the requirement of significant computing resources and high simulation time. Therefore, we have extracted the equivalent Spice model files for the proposed MIV-transistor and the conventional FDSOI transistor without MIV presence. These Spice models are then used to design standard cells for PPA comparison. In this section, we present the strategy for Spice extraction of the FDSOI technology based transistors and in the next section we discuss the simulation results of the standard cell designs with the extracted Spice models.


 We have used the Level 70 (BSIMSOI4 model) Spice parameters in the extraction \cite{hspice_mosfet_models}. Based on our simulation results, BSIMSOI4 model has closely emulated the FDSOI transistor behavior. We have used the Synopsys design flow such as Sentaurus TCAD, to model the device behavior and, extracted the performance characteristics. Later, TCAD2SPICE tool is used to fine-tune the Spice parameters based on the device characteristics. All the simulations done were ran on Intel\textsuperscript{TM} Xeon 2.7GHz CPU with 112 cores and 512 GB RAM.

In the interest of space, the parameter description for level 70 Spice parameters used in the extraction is not thoroughly discussed and a detailed discussion is given in \cite{hspice_mosfet_models}. The nominal values used in extraction are given in Table \ref{tab:Param_constant}. The parameters such as TSI, TOX, TBOX,  L and W  are set to similar process values mentioned in Table \ref{tab:process_spec}. To reduce the complexity in extraction, we have turned-off Flags such as Gate-to-channel current model selector (IGCMOD). The model selector for SOI (SOIMOD) is set to 2, selects an 'ideal FD' scenario. The other model selectors are set to  default selection, and hence are not mentioned in Table \ref{tab:Param_constant}.

\begin{table}[t]
    \centering
        \caption{Level 70 parameters constants and Flags used in extraction}\label{tab:Param_constant}
    \begin{tabular}{|c|p{4cm}|c|c|}
         \hline 
        \textbf{Parameter} &  \textbf{Description} & \textbf{Value} \\
        \hline
        LEVEL & Spice model selector &  70\\
        \hline
        MOBMOD & Mobility model selector &  4 \\
        \hline
        CAPMOD & Flag for the short channel capacitance model &  3 \\
        \hline
        IGCMOD & Gate-to-channel tunneling current model selector &  0 \\
        \hline
        SOIMOD & SOI model selector &  2 \\
        \hline
        TSI & Silicon Thickness (m) &  $7\times10^{-9}$ \\
        \hline 
        TOX & Oxide Thickness  (m) &  $1\times10^{-9}$\\
        \hline
        TBOX & Buried Oxide Thickness  (m) &  $100\times10^{-9}$\\
        \hline
        L & Channel Length (m) &  $48\times10^{-9}$\\
        \hline
        W & Channel Width (m) &  $192\times10^{-9}$\\
        \hline
        
        TNOM & Nominal Temperature (\textsuperscript{o}C) &25\\
        \hline
   \end{tabular}

\end{table}

\subsection{Spice Model Extraction Methodology}\label{subsec:t2s_strategy}

\begin{figure}[]
    \centering
    \includegraphics[width=0.8\linewidth]{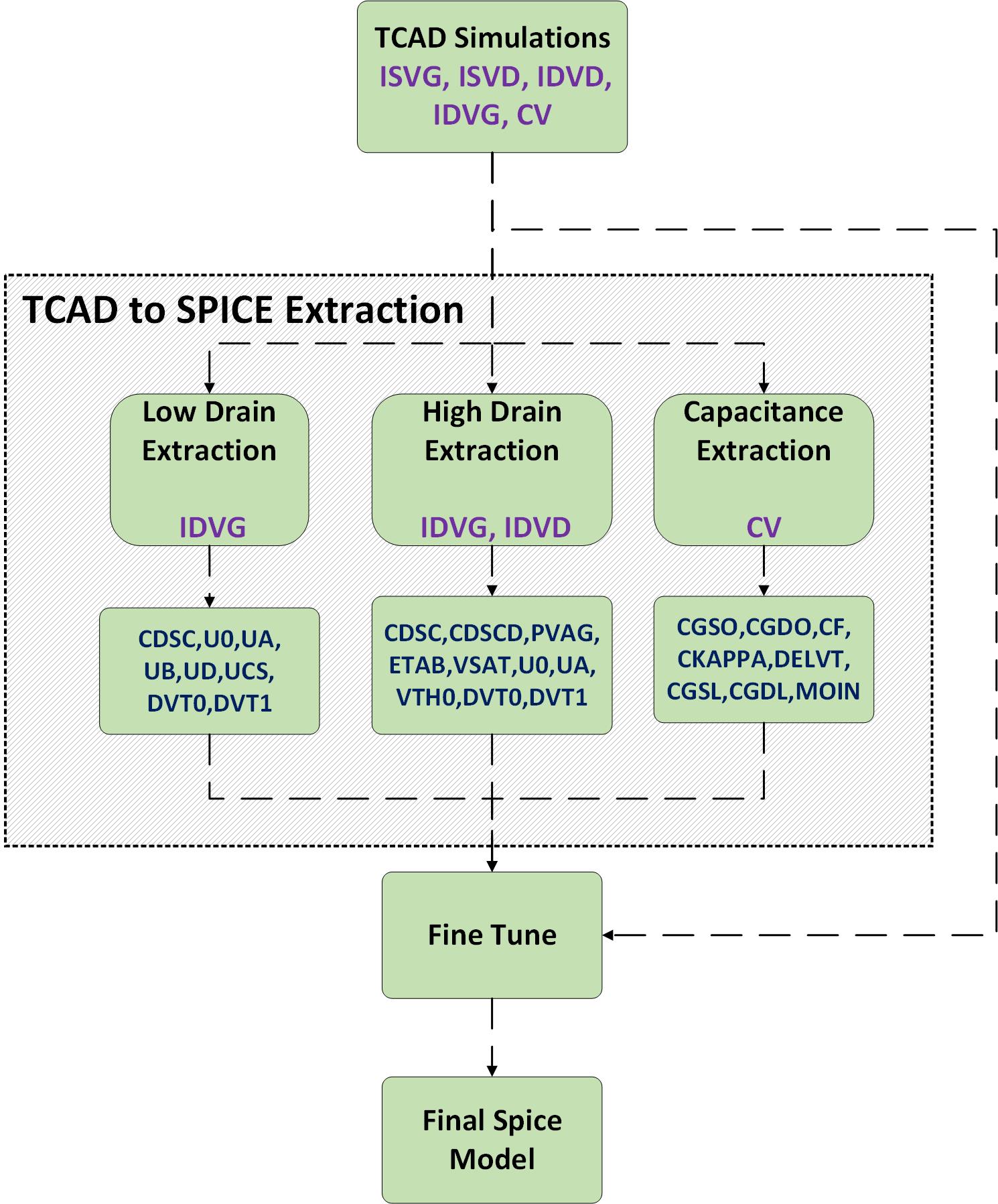}
    \caption{Spice Parameter Extraction}
    \label{fig:spice_extraction}
\end{figure}

The methodology used in Spice extraction is presented in Figure \ref{fig:spice_extraction}. The device characteristics are extracted from the TCAD device simulations for varying voltage biases and are used in fine-tuning the Spice parameters. The extraction was done sequentially:
\begin{enumerate*}
\item Low Drain Extraction, 
\item High Drain Extraction, and 
\item Capacitance Extraction.
\end{enumerate*}
In Low Drain extraction, the device is simulated for lower Drain voltages ($V_{DS}=0.05V$) and device characteristics, such as Source and Drain current with Varying Gate Voltages ($I_{D}\ \&\ I_{S}\ v.s\ V_{G}$) is obtained from the TCAD simulations. Subsequently, for High Drain extraction, higher Drain voltages ($V_{DS}=1.0V$) are used to obtain the $I_{D}\ \&\ I_{S}\ v.s\ V_{G}$ characteristics. Additionally, for High Drain extraction, Source and Drain current with varying Drain voltages with multiple biases ($V_{GS}=0.4V$ to $1.0V$) are used to fine-tune parameters. For  capacitance extraction, Gate capacitance with varying Gate voltages is used to extract the capacitance-related parameters. The methodology compared the device characteristics with the Spice simulations, and the error is used to fine-tune the Spice parameters.  A detailed account for the parameters extracted is presented below:

\begin{enumerate}
    \item For Low Drain extraction, the parameters such as CDSC, U0, UA, UB, UD, UCS, DVT0, and DVT1 are used to extract nominal values for the subsequent extraction. The parameters such as U0, UA, UB, UD, UCS control the mobility of the carrier in Spice models, DVT0 and DVT1 control the short channel effects. The parameters such as U0, UA, DVT0, and DVT1 are passed to the subsequent extraction regions for fine-tuning.
    \item For High Drain extraction, the parameters such as CDSC, CDSCD, U0, UA, VTH0, PVAG, DVT0, DVT1, ETAB, and VSAT are extracted. CDSC, CDSCD and ETAB control the subthreshold slope regions within the $I_{D}\ \&\ I_{S}\ v.s\ V_{G}$ characteristics. VTH0, VSAT and PVAG control the current regions above the threshold voltage. 
    \item For  Capacitance extraction, the parameters such as CKAPPA, DELVT, CF, CGSO, CGDO, MOIN, CGSL, and CGDL are used to model the capacitance behavior. CKAPPA, CGSL, and CGDL control the lower biased regions in the capacitance characteristics. DELVT is used to adjust the threshold voltage. CF, MOIN, CGSO, and CGDO controlled the overall accuracy and, are fine-tuned along with the previous parameters to improve the accuracy.
\end{enumerate}

%% file: Section4b_Extraction_results.tex
\begin{figure*}
\centering
\begin{subfigure}[t]{0.32\linewidth}
\includegraphics[width=0.9\linewidth]{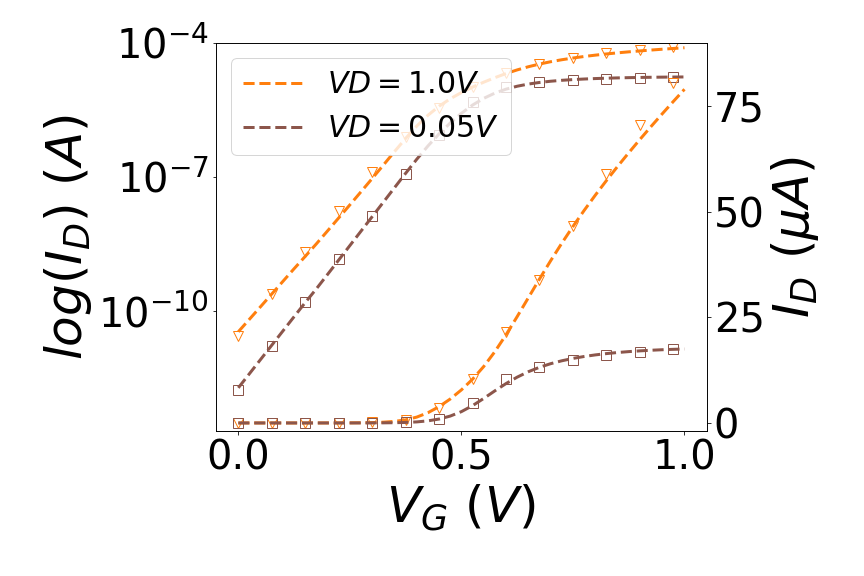}
    \caption{$I_D$ v.s $V_G$ for n-type}
    \label{fig:IDVG_nmos}
\end{subfigure}
\begin{subfigure}[t]{0.32\linewidth}
\includegraphics[width=0.9\linewidth]{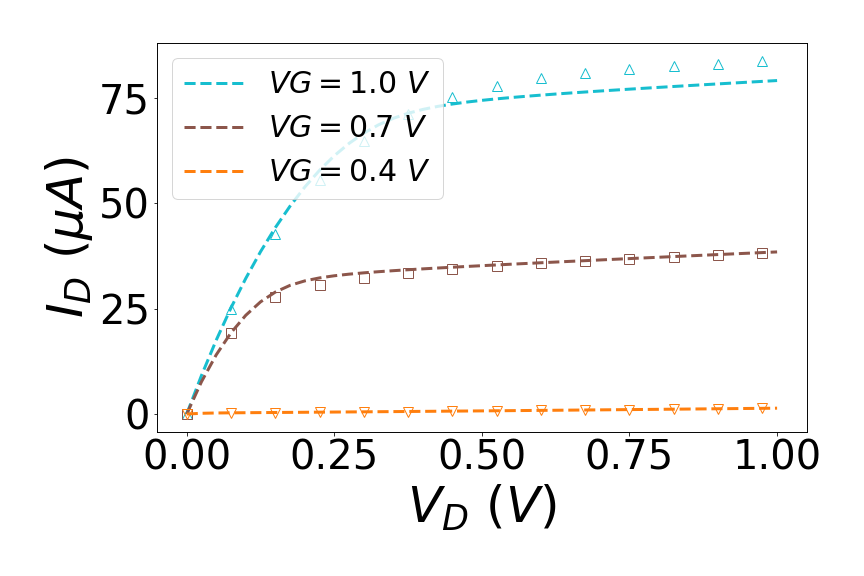}
    \caption{$I_D$ v.s $V_D$ for n-type}
    \label{fig:IDVD_nmos}
\end{subfigure}
\begin{subfigure}[t]{0.32\linewidth}
\includegraphics[width=0.9\linewidth]{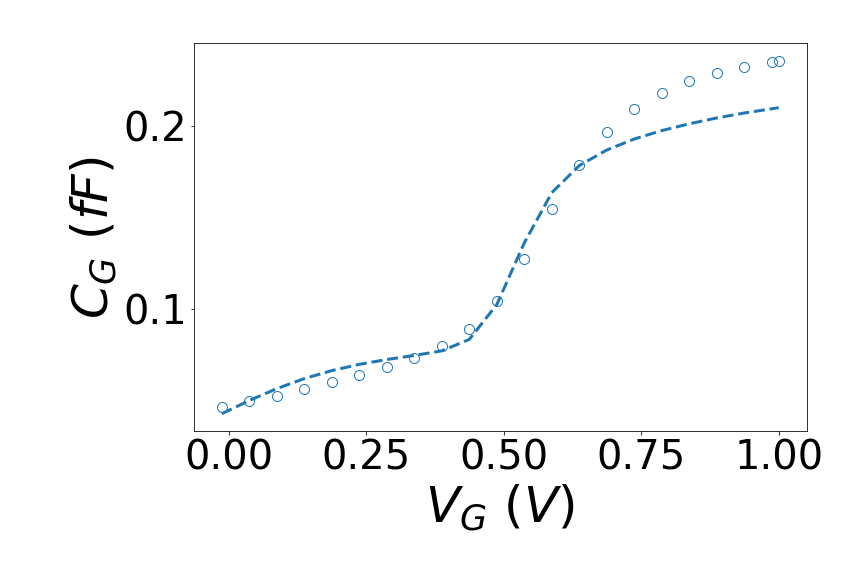}
    \caption{$C_G$ v.s $V_G$ for n-type}
    \label{fig:CV_nmos}
\end{subfigure}
\begin{subfigure}[t]{0.32\linewidth}
\includegraphics[width=0.9\linewidth]{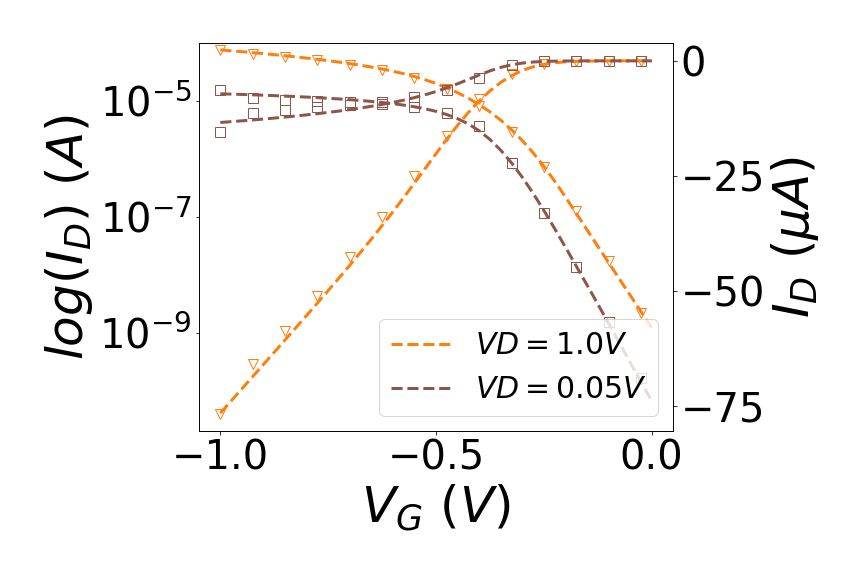}
    \caption{$I_D$ v.s $V_G$ for p-type}
    \label{fig:IDVG_pmos}
\end{subfigure}
\begin{subfigure}[t]{0.32\linewidth}
\includegraphics[width=0.9\linewidth]{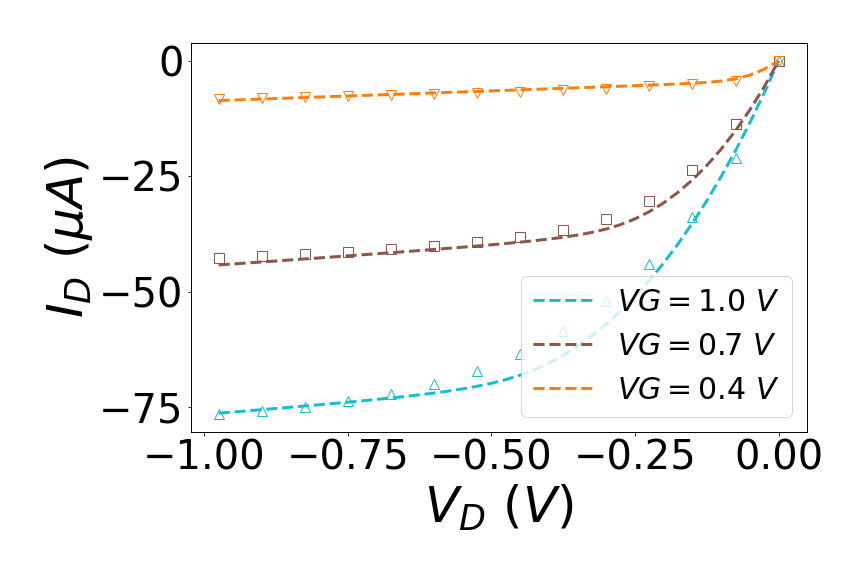}
    \caption{$I_D$ v.s $V_D$ for p-type}
    \label{fig:IDVD_pmos}
\end{subfigure}
\begin{subfigure}[t]{0.32\linewidth}
\includegraphics[width=0.9\linewidth]{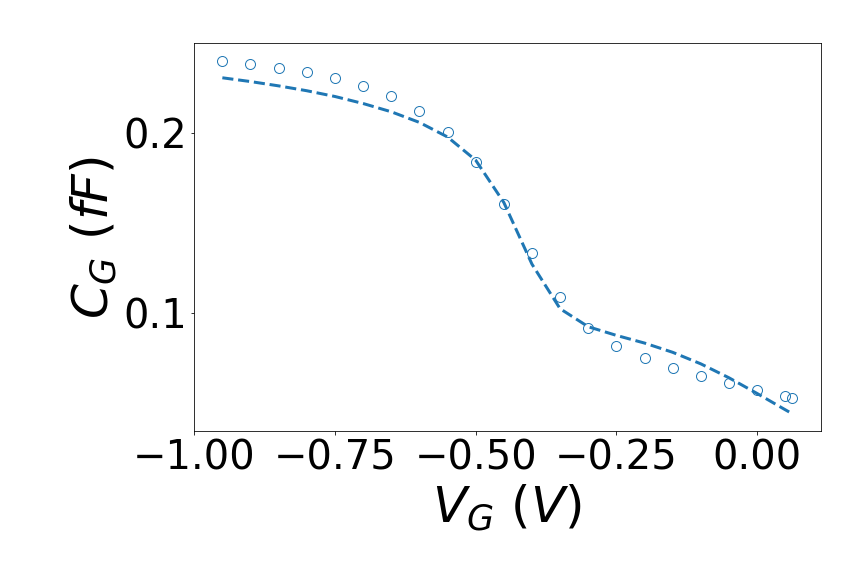}
    \caption{$C_G$ v.s $V_G$ for p-type}
    \label{fig:CV_pmos}
\end{subfigure}
\caption{Level 70 Extraction results for 4-channel MIV-transistor} \label{fig:comparison_for_nominal_case}
\end{figure*}

The overall performance of the spice model extraction results is presented in Table \ref{tab:tcad2spice}. We have compared the transistor characteristics obtained from the TCAD models to the extracted spice model characteristics for n-type and p-type transistors. The process and design parameters of proposed MIV-transistors with 1-channel, 2-channel and 4-channel along with the FDSOI transistor without MIV are given in Table \ref{tab:process_spec}. The overall extraction error was under 10\% for all the cases. The MIV-transistor characteristics using the TCAD models and spice extraction for 4-channel model is presented in Figure \ref{fig:comparison_for_nominal_case}. From the results, we can see that the extracted spice models for the FDSOI transistor type can be used for the design of standard cells and in the next section, we discuss about the Power-Performance-Area (PPA) metric comparison for different standard cell implementations.

\begin{table}
\caption{TCAD to Spice extraction results}\label{tab:tcad2spice}
\centering
    \begin{tabular}{|c|p{0.5cm}|p{0.5cm}|p{0.5cm}|p{0.5cm}|p{0.5cm}|p{0.5cm}|p{0.5cm}|p{0.5cm}|}
    \hline
        \multirow{2}{*}{\textbf{Region}} & \multicolumn{2}{c|}{\textbf{4-channel}} & \multicolumn{2}{c|}{\textbf{2-channel}} & \multicolumn{2}{c|}{\textbf{1-channel}} & \multicolumn{2}{c|}{\textbf{Traditional}} \\
        \cline{2-9}
         & \textbf{ n} & \textbf{ p} & \textbf{ n} & \textbf{ p} & \textbf{ n} & \textbf{ p} & \textbf{ n} & \textbf{ p}\\
        \hline
         \textbf{IDVG} & 7.2\% & 7.1\% & 6.6\% & 7.0\% & 6.4\% & 8.5\% & 7.9\% & 5.5\% \\
         \hline
         \textbf{IDVD} & 3.5\% & 7.2\% & 3.4\% & 6.8\% & 3.2\% & 7.5\% & 3.7\% & 5.2\%
        \\
         \hline
         \textbf{CV} & 7.0\% & 5.7\% & 4.7\% & 6.0\% & 5.0\% & 7.3\% & 9.6\% & 8.6\%
        \\
        \hline
    \end{tabular}
\end{table}

%% file: Section5_Standard_cell_simulation_results.tex
Several basic standarad cells are implemented using the Spice model files extracted from the previous section to investigate the  Power, Performance and Area (PPA) metrics. In the standard cell design, we have assumed the bottom layer and top layer active region to be p-type transistors and n-type transistors respectively. To connect the top and bottom devices, MIVs (both internal and external) are used. We have assumed 2-metal (M1 and M2) interconnect routing layers. Additionally, to facilitate the routing-related parasitics, we have evaluated the resistance of MIV and interconnect to be  $7\Omega$ and $3\Omega$ respectively. To account for parasitic related to Voltage and Ground interconnects, we assumed resistance $5\Omega$ for both cases. The internal interconnect capacitance values such as metal coupling capacitance and fringing capacitance are ignored to limit the complexity of the design. However, we assumed a load capacitance of 1fF for the driving strength. Additionally, as the load capacitance increases the effect of internal RC parasitic reduces significantly on overall power and delay estimation \cite{rc_parasitics_M3DIC}. 

The following standard cells with the assumed parasitic assumptions and the extracted Spice models are designed in Hspice tool: AND2X1, AND3X1, AOI2X1, INV1X1, MUX2X1, NAND2X1, NAND3X1, NOR2X1, NOR3X1, OAI2X1, OR2X1, OR3X1, XNOR2X1 and XOR2X1. The PPA metrics comparison for standard cells such as Delay time, Power, and Area are presented in Figure \ref{fig:PPA_comparison}. For comparison purposes, we have replaced the top n-type transistors of standard cells with the 2D FDSOI transistor similar to the p-type with M1 spacing between MIV and metal layers (This implementation is represented in the figure as 2D, which has two-layer 2D FDSOI based implementation) and proposed n-type MIV-transistors on the top layer with the 2D FDSOI p-type transistor on the bottom layer. The n-type MIV-transistors on the top layers consist of three cases specifically, 1-channel MIV-transistor (1-ch), 2-channel MIV-transistor (2-ch), and 4-channel MIV-transistor (4-ch). A detailed comparison is given below:

\begin{figure}[]
\begin{subfigure}[t]{\linewidth}
\includegraphics[width=\linewidth]{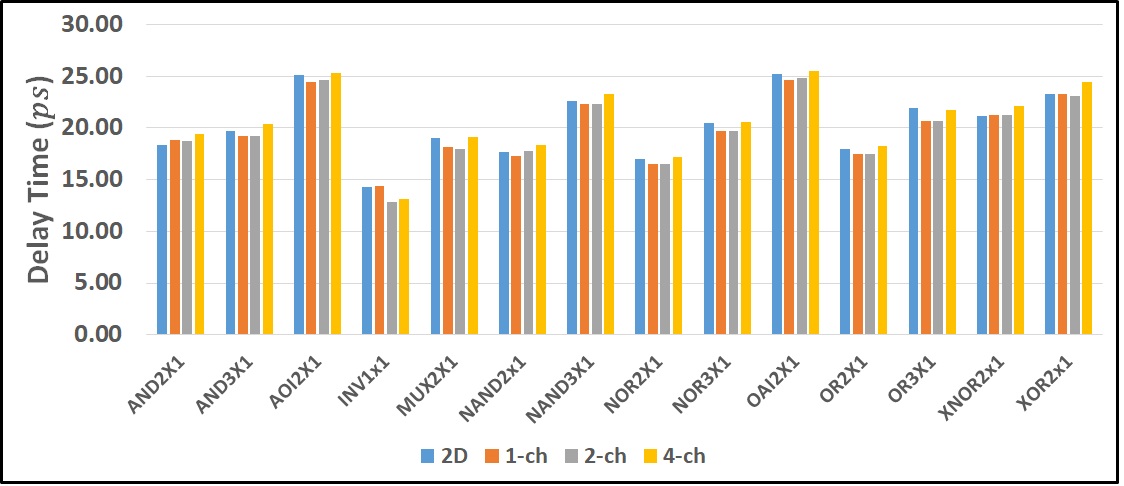}
\caption{}\label{subfig:delay_time_results}
\end{subfigure}
\begin{subfigure}[t]{\linewidth}
\includegraphics[width=\linewidth]{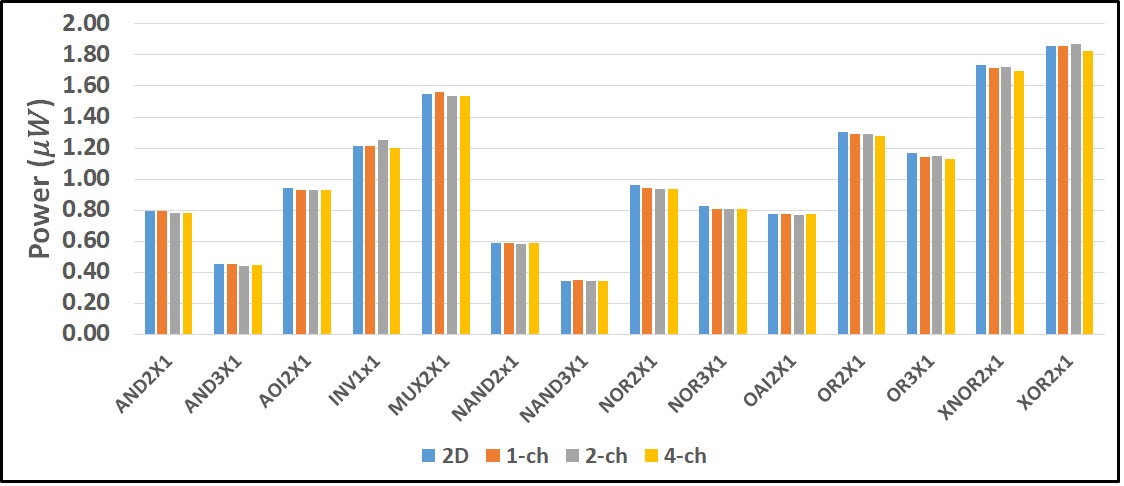}
\caption{}\label{subfig:power_results}
\end{subfigure}
\begin{subfigure}[t]{\linewidth}
\includegraphics[width=\linewidth]{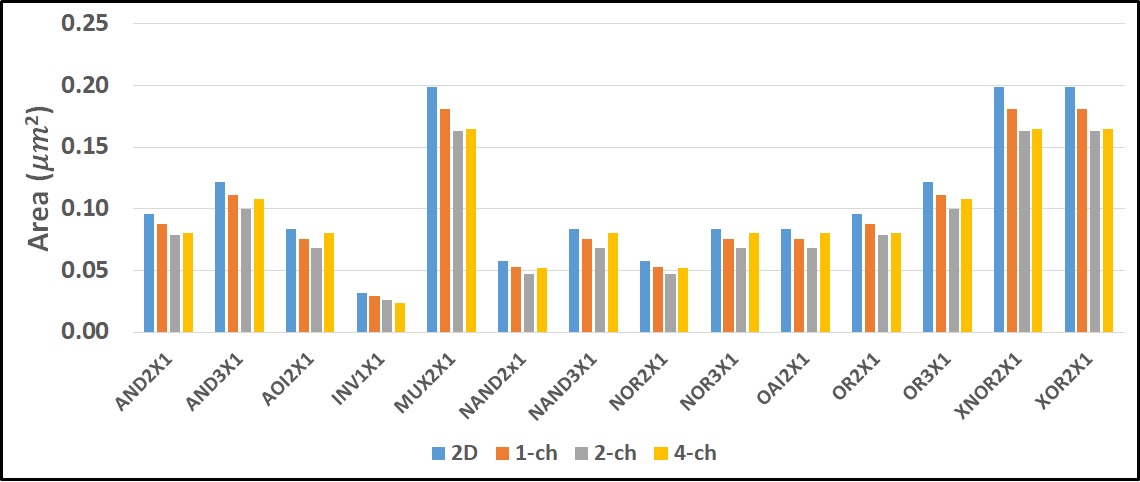}
\caption{}\label{subfig:area_results}
\end{subfigure}
\caption{PPA comparison for the extracted models}\label{fig:PPA_comparison}
\end{figure}


\subsubsection{Delay Time}
The average propagation delay of the outputs from Spice simulations of standard cell designs is presented in Figure \ref{subfig:delay_time_results}. The average delay reduced by  $3\%$, and $2\%$  for  1-channel and 2-channel MIV-transistor based cells respectively. However, the average delay increased by $2\%$  for  4-channel MIV-transistor based cells due to the differences in the transistor characteristics, that is dependent on the model. The AND2X1 cell created using 4-channel model had higher delay time with $6\%$ increase compared with the 2D implementation. The INV1X1 cell created using 2-channel model has delay time reduced by up to $-11\%$ compared with the two-layer 2D FDSOI transistor implementation.

\subsubsection{Power}
The average power consumption of the standard cells from Spice simulations is given in Figure \ref{subfig:power_results}. The average Power reduced by $0.5\%$,  $1\%$, and $2\%$  for  1-channel, 2-channel, 4-channel MIV-transistor based cells respectively. The INV1X1 cell created using the 2-channel model had higher power with $3\%$ increase compared with the two-layer 2D FDSOI transistor-based implementation. The OR3X1 cell created using the 4-channel model has reduced power by up to $3\%$ compared with the two-layer 2D FDSOI transistor-based design. The differences in the power are also accounted due to the differences in the transistor characteristics of the MIV-transistor models.   

\subsubsection{Layout Area}
 For layout area comparison, we have assumed the M1 metal layer separation specifically $24 nm$. Therefore, the layout area on the top layer devices considers the MIV with the M1 metal layer separation as the keep-out-zone. The total layout area of the standard cells are presented in Figure \ref{subfig:area_results} calculated similarly to \cite{mono3D}. Note that this layout area is obtained by considering the maximum layout dimensions on both top-layer and bottom-layer implementation so that the standard cell placement treats both n-type and p-type device layers together. From the figure, the layout area of the standard cells is reduced by upto $9\%$, $18\%$, and $12\%$ on average for 1-channel, 2-channel, and 4-channel MIV-transistor models compared with the standard two-layer FDSOI implementation respectively. 

In addition, we can reduce the total substrate area consumption which is the sum of the bottom layer p-type FDSOI transistor area, and the top layer n-type transistor area considering the MIV placement by up to 31\%. However, this requires separate placement algorithms to optimize the layout area for both layers separately, and also considers the delay and routing resources. In the future, we plan to investigate the placement algorithms that consider the bottom-layer and top-layer device placement separately.


In summary, the proposed MIV-transistor-based FDSOI models provide alternative design choices where PPA metrics can be leveraged. For example, if the delay can be leveraged and the area is limited, the 4-channel MIV-transistor-based standard cells can reduce the area consumption by $25\%$. If the designer is looking for overall improvement, out of all standard cells presented in the paper, the 2-channel model had a $3\%$  reduction in the  average power delay product, and 18\% overall area reduction.

%% file: Section6_Conclusions.tex
This paper studies the design of MIV-transistor in 2-layer M3D-IC using FDSOI process. We looked into the design of MIV-transistor which utilizes the area around MIV to create transistors. For performance metrics comparison, we extracted the spice models of the proposed transistors using the level 70 spice parameters. 14 Gate models have been created to compare the performance using different MIV-transistor models.  Simulation results from standard cell designs suggest that the proposed methodology can reduce 18\% layout area on average compared to the traditional approach. In addition, power consumption and delay time of the standard cells are reduced by 1\% and 3\% on average respectively.
